\documentstyle{mn}

\input{epsf}

\begin{document}

\title[Modelling the evolution of galaxy clustering]
{Modelling the evolution of galaxy clustering}

\author[Baugh \etal]
{C.M. Baugh$^{1}$, A.J. Benson$^{1}$, S. Cole$^{1}$, 
C.S. Frenk$^{1}$ and C.G. Lacey$^{2}$\\
1. University of Durham, 
Department of Physics, Science Laboratories, South Road, Durham DH1 3LE \\
2. Theoretical Astrophysics Centre, Juliane Maries Vej 30, 
Copenhagen, Denmark.\\
Journal Ref: 1999, MNRAS, 305, L21-L25 
}

\maketitle 

\def\mpc {h^{-1} {\rm{Mpc}}}
\def\etal  {\it {et al.} \rm}
\def\rmd {\rm d}
\def\gsim{ \lower .75ex \hbox{$\sim$} \llap{\raise .27ex \hbox{$>$}}}
\def\lsim{ \lower .75ex \hbox{$\sim$} \llap{\raise .27ex \hbox{$<$}}}

\begin{abstract}
Measurements of galaxy clustering are now becoming possible over a range of
redshifts out to $z \sim 3$.  We use a semi-analytic model of galaxy
formation to compute the expected evolution of the galaxy correlation
function with redshift. We illustrate how the degree of clustering
evolution is sensitive to the details of sample selection.  For a fixed
apparent magnitude limit, galaxies selected at higher redshifts are located
in progressively rarer dark matter haloes, compared with the general
population of galaxies in place at each redshift.  As a result these
galaxies are highly {\it biased} tracers of the underlying dark matter
distribution and exhibit stronger clustering than the dark matter.  In
general, the correlation length measured in comoving units, decreases at
first with increasing redshift, before increasing again at higher
redshift. We show that the $\epsilon$-model often used to interpret the
angular correlation function of faint galaxies gives an inadequate
description of the evolution of clustering, and offers no physical insight
into the clustering process.  We compare our predictions with those of a
simple, popular model in which a one-to-one correspondence between
galaxies and dark halos is assumed. Qualitatively, this model reproduces
the correct evolutionary behaviour at high redshift, but the quantitative
results can be significantly in error. Our theoretical expectations are in
good agreement with the high redshift clustering data of Carlberg \etal and
Postman \etal but are higher than the measurements of Le F\`evre \etal

\end{abstract}


\section{Introduction}

In general, the clustering properties of galaxies are expected to vary with
cosmic time. The nature of this evolution depends on factors such
as the values of the cosmological parameters, the identity of the dark
matter and the details of the galaxy formation process.  Traditionally,
attempts to infer the evolution of galaxy clustering have relied on
measurements of the angular correlation function of faint
galaxies, $\omega(\theta)$, as a function of apparent magnitude
(some recent examples include Efstathiou \etal 1991; 
Roche \etal 1993;
Infante \& Pritchet 1995; 
Hudon \& Lilly 1996; 
Brainerd \& Smail  1998; 
Postman \etal 1998).  
The spatial correlation function, $\xi(r)$, may be inferred from 
the diluted signal in projection by inverting Limber's (1954) 
equation given assumptions about the redshift distribution of 
the galaxies and the evolution of their clustering. A simple
model of the evolution of the two-point spatial correlation function, 
the `$\epsilon$-model' (Groth \& Peebles 1977), has often been used to 
interpret this kind of data (Efstathiou \etal 1991). 
Recently, it has become possible to measure the 
spatial correlations directly in deep spectroscopic
surveys (Le F\`{e}vre \etal 1996; 
Giavalisco \etal 1998; Carlberg \etal 1999).

The main difficulty in interpreting measurements of galaxy clustering is
the uncertain relation between the observed galaxies and the underlying
mass distribution. This problem is compounded by the fact
that the properties of the galaxies selected according to any simple
criteria (eg a magnitude limit) are likely to vary over a long redshift
baseline. Thus, the interpretation of the data requires a model for galaxy
evolution. In this paper we use a semi-analytic model of galaxy formation
in which the observable properties of galaxies are calculated {\it ab
initio} within a specified cosmological model.
The semi-analytic model that we use is a development of the one described
in a series of earlier papers (Cole \etal 1994; Heyl \etal 1995; Baugh,
Cole \& Frenk 1996a,b) and is fully discussed in Cole \etal (in preparation). 
Models of this kind have been successful in accounting for many 
properties of the galaxy population such as the shape of the luminosity 
function, the distribution of colours, and the faint counts 
(White \& Frenk 1991; Lacey\etal 1993; Kauffmann, White \& Guiderdoni 1993; 
Cole \etal 1994; Somerville and Primack 1998).  

Semi-analytic models have been used to investigate the clustering of
galaxies (e.g. Kauffmann, Nusser \& Steinmetz 1997; Kauffmann \etal
1998, 1999). An earlier version of the model employed here was used
to predict that the `Lyman-break' galaxies discovered by Steidel \etal
(1996) at $z \sim 3$ should have a comoving correlation length similar to
that of bright galaxies at the present day (Baugh \etal 1998; see also
Davis \etal 1985).  This theoretical prediction was subsequently confirmed
by observations (Adelberger \etal 1998; Giavalisco \etal 1998).  Baugh
\etal (1998) predicted that Lyman break galaxies are hosted by the most
massive and therefore the rarest dark matter haloes in place at $z
\sim 3$.  Such haloes are biased tracers of the underlying mass 
distribution and are more strongly clustered than the dark matter
which, in current models of structure formation, was much more weakly
clustered at $z\sim3$ than at present.

\section{Modelling galaxy clustering}

\subsection{The $\epsilon$-model for the two-point galaxy correlation
function}

The correlation function of local, optically selected galaxies is 
well described by a simple power law $\xi(r) = (r_{0}/r)^{\gamma}$, 
with a slope of $\gamma \approx 1.8$ and a correlation length 
of $r_{0} \approx 5 h^{-1}$Mpc (Groth \& Peebles 1977) 
for $r < 10 h^{-1}$Mpc. 
Groth \& Peebles (1977) proposed the
``$\epsilon$-model" to describe the redshift evolution of the correlation
function, measured in terms of proper separation:

\begin{equation}
\xi(r,z) = \left( \frac{r_{0}}{r} \right)^{\gamma} (1 + z)^{-(3+\epsilon)}.
\end{equation} 
If $\epsilon = 0$, the product of the proper number density of galaxies
and the correlation function is constant and the clustering pattern is
fixed in proper coordinates; this is known as {\it stable} clustering
(Peebles 1980, \S 56). The correlation length, $x_{0}$, can also be
written in comoving units as

\begin{equation}
x_{0} = r_{0} (1+z)^{-(3+\epsilon-\gamma)/\gamma},
\end{equation}
where $r_{0}=x_{0}(z=0)$.
The clustering pattern is fixed in comoving coordinates if 
$\epsilon = \gamma-3 \approx -1.2$, whilst the correlation 
function evolves as expected in linear perturbation theory, 
for $\Omega=1$, if $\epsilon = \gamma - 1 \approx 0.8$.

\subsection{The hierarchical clustering of dark matter}

Significant progress has been made in our understanding of the dynamical
evolution of dark matter in the universe, primarily as a result of N-body
simulations (e.g. Jenkins \etal 1998). The growth of clustering in the
dark matter can be approximated by a scaling formula that transforms 
the amplitude and scale of a linear density fluctuation into the 
corresponding values for a nonlinear fluctuation (Hamilton \etal 1991;  
Peacock \& Dodds 1996).

The clustering evolution of dark matter haloes is significantly
different from that of the underlying dark matter (e.g. Cole \& Kaiser
1989; Mo \& White 1996; Kauffmann, Nusser \& Steinmetz 1997; Bagla 1998;
Matarrese \etal 1997; Moscardini \etal 1998).
Haloes more massive than the characteristic mass $M_{*}(z)$
\footnote 
{The characteristic mass is defined by $\sigma(M_{*})=\delta_c(z)$, 
where $\sigma(M)$ is the {\it rms} linear density fluctuation 
at z=0, and $\delta_c(z)$ is the extrapolated critical linear 
overdensity for collapse at redshift $z$ -- see Lacey \& Cole 1993.}
initially display stronger clustering than the dark matter.
The clustering of the dark matter subsequently grows 
faster than that of these haloes, 
reducing their bias.
The bias is further diluted by the subsequent formation of haloes 
of mass $M$, when $M \le M_*$. 
The clustering evolution of dark haloes sets the scene for understanding 
the clustering evolution of galaxies. 

\subsection{Galaxy clustering in a semi-analytic model}

The expected number of galaxies per dark matter halo and its associated
scatter depend on the details of the galaxy formation process (Benson
\etal 1999), while the observed number per halo depends also on the 
selection criteria.  
Semi-analytic modelling provides the means to calculate how dark haloes are 
populated by visible galaxies and also to derive the expected galaxy
correlation function. The latter requires the following four steps:

\begin{itemize} 
\item[(i)] Obtain the nonlinear correlation function of the  
dark matter distribution using the scaling formula of 
Peacock \& Dodds (1996).

\item[(ii)] Select model galaxies according to the observational
selection criteria.

\item[(iii)] Compute an effective bias parameter, $b_{{\rm eff}}$ (assumed
to be independent of scale) using equation~(\ref{eq:bias1})
or~(\ref{eq:bias2}) below, weighting each halo by the number of 
galaxies that satisfy the observational selection criteria:
\begin{equation}
b_{\rm eff} = 
\frac{
\int b(M) N_{gal}(M) n(M) {\rm d} M
} 
{
\int N_{gal}(M) n(M) {\rm d} M
},
\label{eq:bias}
\end{equation}
where, $b(M)$ is the bias parameter
of dark matter haloes of mass $M$; 
$N_{gal}(M)$ denotes the mean number of galaxies in a halo of mass
$M$ that satisfy the selection criteria; and $n(M)$ is the number 
density of dark matter haloes of mass $M$.

\item[(iv)] Obtain the real space correlation function of galaxies
(ie. free of distortions due to peculiar velocities) by multiplying the
nonlinear dark matter correlation function by the effective bias parameter
squared, $\xi_{gal} = b^{2}_{{\rm eff}} \xi_{\rm DM}$.
\end{itemize}

The bias parameter for dark matter haloes of mass $M$ is given by 
(Mo \& White 1996 ; see also Cole \& Kaiser 1989):

\begin{equation}
b_{1}(M, z) = 1 + \frac{1}{\delta_c(z)D(z)} \left[
\left(\frac{\delta_c(z)}{\sigma(M)}\right)^2 -1 \right], 
\label{eq:bias1}
\end{equation}
where $D(z)$ is the linear growth factor, normalized to unity at the
present day. 
This expression has been tested against N-body
simulations (Mo \& White 1996; Mo, Jing \& White 1996) and 
works in practice down to the Lagrangian radius of the dark matter
halo, $r \approx r_{L}$, where $r_{L} = \left(3M/4\pi
\rho_{0}\right)^{1/3}$ and $\rho_{0}$ is the present mean density, even in
the mildly nonlinear regime where $\xi(r) \sim 1$.  
Jing (1998) extended the comparison to higher resolution simulations 
and found that equation~(\ref{eq:bias1}) systematically underpredicts 
the bias of haloes with $M/M_{*} \ll 1$; better agreement 
is obtained with the slightly modified prescription:
\begin{equation}
b_{2}(M,z) = b_{1}(M,z) 
\left( \frac{0.5}{(\delta_c(z)/\sigma(M))^{4}} + 1 \right)^{(0.06 - 0.02n)},
\label{eq:bias2}
\end{equation}
where $n$ is the effective spectral index of the power spectrum, 
${\rm d}\ln P(k)/{\rm d}\ln k$,  at the wavenumber defined by the 
Lagrangian radius of the dark matter halo, $k = 2 \pi /r_{L}$.

\section{Results}

\begin{figure}
{\epsfxsize=8.truecm \epsfysize=10.truecm 
\epsfbox[30 190 375 660]{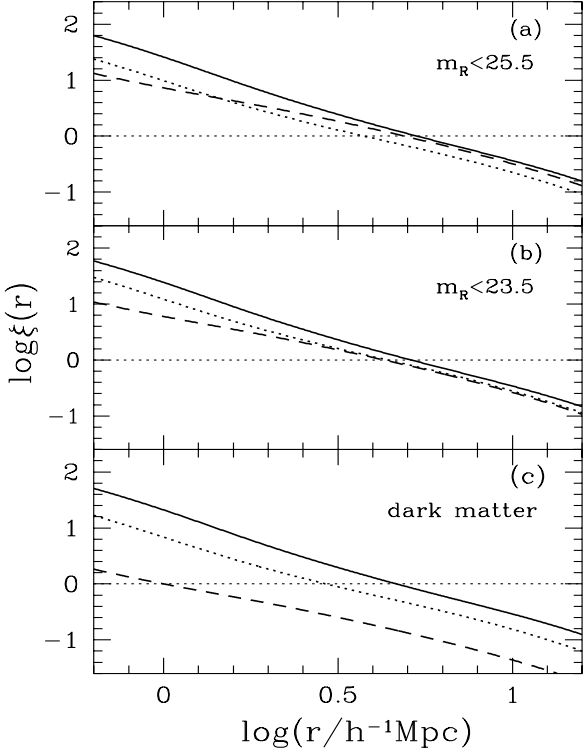}}
\caption[]
{
The correlation function of galaxies brighter than (a) $m_{R}= 25.5$ and
(b) $m_{R} = 23.5$, as a function of redshift, plotted against comoving
separation.  
The correlation function of the dark matter is shown in (c).
The correlation functions are plotted at $z=0.25$ (solid line), 
$z=1$ (dotted) and $z=3$ (dashed).
The horizontal dotted line marks $\xi(x_{0})=1$ which defines the 
comoving correlation length, $x_{0}$.
}
\label{fig:1}
\end{figure}

\begin{figure}
{\epsfxsize=8.truecm \epsfysize=10.truecm 
\epsfbox[80 160 550 660]{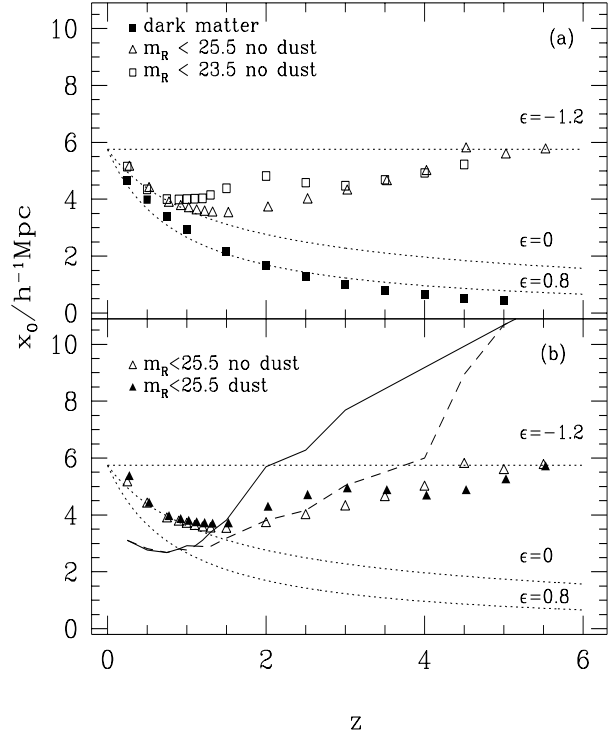}}
\caption[]
{
The comoving correlation length, defined as the separation for
which $\xi(x_{0})=1$, plotted as a function of redshift. 
(a). The correlation length of galaxies brighter than $m_{R} =
23.5$ (open squares) and that of galaxies brighter than $m_{R} = 25.5$ 
(open triangles). 
The filled squares show the comoving correlation length of the dark matter.  
The dotted lines show the evolution expected in the $\epsilon$-model 
of eqn~2, with $\epsilon=-1.2, 0$ and $0.8$, 
assuming $\gamma=1.8$ and $x_{0}(z=0) = 5.75 h^{-1}{\rm Mpc}$. 
(b). The effects of dust are demonstrated by the filled
triangles for galaxies brighter than $m_R=25.5$. 
The dashed line shows the correlation length predicted by a model that 
assumes a one-to-one correspondence between galaxies and 
dark halos down to a lower halo mass limit, chosen such that the 
number density of galaxies in the semi-analytic model, without dust, 
is reproduced. The solid line shows the result when dust is included.
}
\label{fig:2}
\end{figure}

\begin{figure}
{\epsfxsize=8.truecm \epsfysize=10.truecm 
\epsfbox[60 180 460 660]{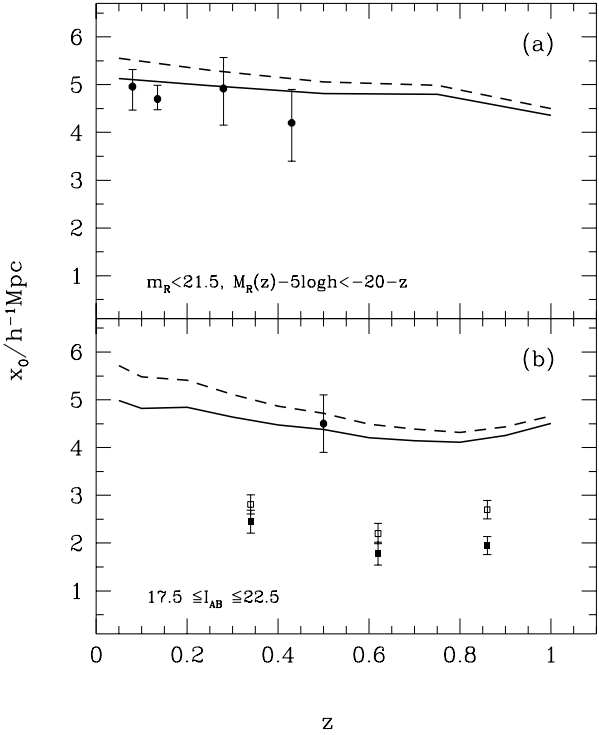}}
\caption[]
{
The evolution of the comoving correlation length of galaxies out to $z \sim
1$. The lines show the predictions of the semi-analytic model, using the
bias parameter given by eqn.~(\ref{eq:bias1}) (solid line) and by
eqn.~(\ref{eq:bias2}) (dashed line.) 
In (a), galaxies are selected with apparent magnitude $m_R < 21.5$ 
and absolute magnitude $M_{R} < -20-z$.  
The points show a preliminary analysis of clustering in the CNOC2 field
survey (Carlberg \etal 1999).  
In (b), galaxies are selected in the observer frame $I$-band with 
$17.5 \le I_{AB} \le 22.5$.
The squares show the correlation length in comoving
units derived from the CFRS by LeF\`{e}vre \etal (1996) for $\Omega=1$
(filled) and $\Omega=0$ (open).  
The filled circle shows the correlation length inferred 
from the angular clustering over a larger area by Postman \etal (1998).
}
\label{fig:3}
\end{figure}

We consider the clustering of galaxies in a low density, flat  
cold dark matter universe, with present-day density parameter 
$\Omega_{0} = 0.3$, and cosmological constant $\Lambda_{0} = 0.7$, 
normalised to reproduce the observed abundance of rich clusters 
($\sigma_8=0.94$; Eke, Cole \& Frenk 1996). 
Benson \etal (1999) show that the correlation function of 
$L>L_{*}$ galaxies in this model is in good agreement 
with that measured for APM Survey galaxies (Baugh 1996) over 
separations in the range $0.3 h^{-1} {\rm Mpc} < r < 15 h^{-1} {\rm Mpc}$. 
This agreement is not reproduced in critical density models.

The expected evolution of galaxy clustering for 
samples selected at magnitude limits of $m_R = 25.5$ and 
$m_R = 23.5$ is shown in Figs~\ref{fig:1}a and~\ref{fig:1}b 
respectively. The correlation functions are plotted at redshifts of  
$z=0.25$ (solid line), $1.0$ (dotted line) and $3.0$ (dashed line). 
The corresponding dark matter correlation functions are shown in
Fig~\ref{fig:1}c. 
It is clear from Fig.~1 that the correlation function of galaxies changes
little over the entire redshift range $0<z<3$. 
The correlation function of the mass, on the other hand,
grows appreciably between $z=3$ ($x_{0}=1 h^{-1}{\rm Mpc}$) and $z=0.25$
($x_{0}=4.6h^{-1}{\rm Mpc}$). The markedly different evolutionary rates
of the clustering of galaxies and mass was already noted in the first
simulations of biased galaxy formation in a cold dark matter universe
(Davis \etal 1985).

The non-monotonic variation of the correlation length (in comoving units)
with redshift is shown in Fig.~\ref{fig:2}. The top panel displays 
the evolution of the comoving correlation length of galaxies with 
apparent magnitude brighter than $m_R=23.5$ (open squares) 
and $m_R=25.5$ (open triangles). 
Only points for which there are fifty or more model galaxies 
in the sample are plotted.
The internal errors in our estimate of the correlation length are typically 
smaller than the symbol size, except for the point at $z=4.5$ for galaxies
brighter than $m_R <23.5$; the uncertainty in this point makes it consistent 
with the correlation length obtained for the fainter sample.
The correlation length typically decreases at first with 
increasing lookback time, reaches a minimum around redshift 1--2 and 
then grows again as the selected galaxies find themselves 
in increasingly rare haloes. 
Similar behaviour was seen in a simulation of the same cosmology 
by Kauffmann \etal (1998). 
The number density of galaxies in apparent magnitude selected samples 
falls rapidly with increasing redshift; at the same time, the bias parameter 
of the occupied haloes is increasing. The interplay between these effects 
can give rise to features in the comoving correlation length, such as the 
small peak at $z \sim 2$ in the $m_R < 23.5$ sample.

The evolution of the galaxy clustering length is very different 
from that of the dark matter, shown by the filled squares in 
Fig.~\ref{fig:2}(a). 
For the cases illustrated here, the comoving 
correlation length of galaxies at $z=4$ lies in the 
range $x_{0}=5$--$5.5 h^{-1}{\rm Mpc}$, whereas $x_{0} = 0.5h^{-1}$Mpc 
for the dark matter.
The dotted lines in Fig.~\ref{fig:2} show the $\epsilon$-model
for various values of the parameter $\epsilon$. Evidently, there is no
single value of this parameter that gives an acceptable
match to the galaxy data over the whole redshift range, 
a conclusion suggested by other workers using different 
techniques (Giavalisco \etal 1998; Connolly, Szalay \&
Brunner 1998; Matarrese \etal 1997; Moscardini \etal 1998).  

In Fig.\ref{fig:2}(b), we show how the evolution of the correlation
length is altered when the effects of obscuration by dust are included (see
Cole \etal in preparation).  For redshifts
$z>2$, the correlation length shows less evolution when dust is included,
though the differences are relatively minor. 

The dashed and solid lines in Fig.\ref{fig:2}(b) show the predictions of a
simple clustering model which is often used in the absence of a physical
description of galaxy formation. 
In this model, a one-to-one correspondence between galaxy luminosity 
and the mass of the host dark matter halo is assumed.
Haloes are populated with single galaxies down to some minimum mass, 
chosen in order to match an observed galaxy number density
(see for example Moscardini \etal 1998; Mo, Mao \& White 1999; 
Adelberger etal 1998).
In Fig.~\ref{fig:2}(b), the minimum halo mass at each redshift is set  
in order to match the number density of galaxies in the semi-analytic model. 
The solid line shows the results of applying the simple model to match 
the number density in the semi-analytic model with dust; the dashed line 
shows the case when dust is neglected. 
Note that the predicted number densities of galaxies are different in 
these two cases for $z>1$. 
The model with dust gives one galaxy per 
square arcminute that satisfy the colour selection used 
to isolate `Lyman-break' galaxies, in agreement with the observed 
value (Adelberger \etal 1998). 

The ``one-galaxy-per-halo" model displays qualitatively similar 
clustering evolution to that predicted by the semi-analytic model.  
However, quantitatively, the simple model seriously underpredicts 
the correlation length at low redshifts, where haloes that host more 
than one galaxy make a significant contribution to the clustering.  
At high redshifts, where one might at first expect the
simple model to be more appropriate, the clustering length can be up to a
factor of two greater than in the semi-analytic model, when the effects 
of dust are included. 
In the semi-analytic model, we find that the mean UV luminosity of 
the central galaxy in a dark matter halo increases with halo mass, 
though there is scatter in this relationship. 
This means that there is {\it not} a sharp transition in mass between  
haloes that never contain galaxies bright enough to be included in 
the sample and haloes for which the mean galaxy occupation number is 
unity. 
In addition at high redshift, the host haloes are more massive 
than $M_*$, so their abundance falls exponentially with 
increasing mass.
The interplay between these effects means that haloes for which 
the mean galaxy occupation number is less than unity can make a 
significant contribution to the clustering amplitude.
Assuming that there is exactly one galaxy per halo and that 
the dark matter halo abundance matches a given galaxy abundance, results 
in too large a dark halo mass being assigned to each galaxy;  
consequently the ``one-galaxy-per-halo'' model overestimates   
the correlation strength.

We compare the predicted clustering of galaxies in our model with
observational data out to $z \sim 1$ in Fig.~\ref{fig:3}. In the top panel,
the data points come from a preliminary analysis of clustering in the CNOC2
field survey by Carlberg \etal (1999). These authors consider
galaxies brighter than the spectroscopic limit of their survey,
$m_{R}=21.5$, and use the measured redshift of each galaxy to estimate an
absolute magnitude, $M$. Galaxies brighter than $M=-20-z$ are then
selected, where the redshift dependence represents an approximate 
correction for band shifting and evolution.
We apply the same selection criteria to galaxies in our
semi-analytic model. The solid line in Fig.~\ref{fig:3}. shows the
predicted correlation length obtained using equation~(\ref{eq:bias1}), while
the dashed line shows the predictions using the modified expression for
the bias given by equation~(\ref{eq:bias2}). 
Fig.~\ref{fig:3}(b) shows the model predictions for galaxies 
with observer frame $I$-band magnitudes in the 
range $17.5 \le I_{AB} \le 22.5$. The squares show the
comoving correlation lengths estimated from the Canada-France Redshift
Survey (CFRS) by LeF\`{e}vre \etal (1996), for two different assumptions
for the underlying cosmology. The solid circle comes from Postman
\etal (1998) who analysed angular correlations over a much larger solid
angle, applying a similar magnitude selection to that in the CFRS. The
correlation length they obtain is a factor of 2 larger than the CFRS value.
Postman \etal argue that the CFRS fields are too small and that sampling
fluctuations of this magnitude are expected in fields of this size. Our model
predictions agree very well with the results of Postman \etal

\section{Discussion and conclusions}
We have shown that the application of simple selection criteria 
(e.g. an apparent magnitude limit) can lead to a {\it non-monotonic} 
dependence of clustering strength on redshift.
The often used ``one-galaxy-per-halo" model predicts clustering 
evolution that is qualitatively similar to that predicted by the 
semi-analytic model. However this simple model can give
results that are quite wrong in detail. The relatively complex behaviour
of the galaxy clustering strength with redshift is not reproduced by the
$\epsilon$-model which, furthermore, offers no insight into the 
physical processes that drive the evolution of clustering.

Our results and those of Benson \etal (1999) indicate that a 
cold dark matter model that agrees well with galaxy  clustering in 
the local universe also agrees well with clustering data at high redshift. 
In particular our $\Omega_0=0.3$, $\Lambda_{0}=0.7$, $\sigma_8=0.94$ 
cold dark matter model matches the APM correlation function at low 
redshift (Benson \etal 1999), the CNOC2 (Carlberg \etal 1999) 
and Postman \etal (1998) data at $z\lsim 0.5$ and 
the Lyman-break galaxy data (Adelberger \etal 1998) at $z\simeq 3$ 
(Baugh \etal 1998; Governato \etal 1998).

There is little doubt that forthcoming large-area imaging surveys and
catalogues of galaxy redshifts will lead to an explosion in the 
quantity of data bearing on the evolution of galaxy clustering. 
It is unclear whether these data will constrain cosmological parameters 
or the nature of galaxy evolution. 
In either case, the modelling techniques used in this paper, and those used in
related studies (Governato \etal 1998; Benson \etal 1999; Kauffmann
\etal 1998, 1999), 
will play an important role in the interpretation of the data.

\section*{Acknowledgements} 
AJB, SMC and CSF acknowledge receipt of a PPARC Studentship, Advanced
Fellowship and Senior Fellowship respectively. CGL was supported by the
Danish National Research Foundation through its establishment of the
Theoretical Astrophysics Center. This work was supported in part by a
PPARC rolling grant, the EC TMR Network on The Formation and Evolution 
of Galaxies and by a Durham University computer equipment grant.

\setlength{\parindent}{0cm}
\vspace{0.2cm}
{\bf References} 

\small

\newcommand{\mn}{{ MNRAS, }}
\newcommand{\apj}{{ ApJ, }}
\newcommand{\apjs}{{ ApJS, }}
\newcommand{\aj}{{ AJ, }}
\renewcommand{\aa}{{ AA, }}
\newcommand{\nat}{{ Nature}}
\def\refe {\par \hangindent=.7cm \hangafter=1 \noindent}

\refe Adelberger, K., Steidel, C.C., Giavalisco, M., Dickinson, M., 
      Pettini, M., Kellogg, M., 1998, \apj 505, 18
\refe Bagla, J.S., 1998, \mn 299, 417
\refe Baugh, C.M., 1996, \mn 280, 267
\refe Baugh, C.M., Cole, S., Frenk, C.S., 1996a, \mn 283, 1361
\refe Baugh, C.M., Cole, S., Frenk, C.S., 1996b, \mn 282, L27
\refe Baugh, C.M., Cole, S., Frenk, C.S., Lacey, C.G., 1998, \apj 498, 504
\refe Benson, A.J., \etal 1999, MNRAS submitted, astro-ph/9903343 
\refe Brainerd, T.G., Smail, I., 1998, \apj 494, L137
\refe Carlberg, R.G.,  \etal 1999, Proc. Roy. Soc. London, 357, 167
\refe Cole, S., Arag\'{o}n-Salamanca, A., Frenk, C.S., 
      Navarro, J.F., Zepf, S.E., 1994, \mn\, 271, 781
\refe Cole, S., Lacey, C.G., Baugh, C.M., Frenk, C.S., 1999, in preparation
\refe Cole, S., Kaiser, N., 1989, \mn 237, 1127
\refe Connolly, A.J., Szalay, A.S., Brunner, R.J., 1998, \apj 499, L125
\refe Davis, M., Efstathiou, G., Frenk, C.S., White, S.D.M., 1985, 
      \apj 292, 371
\refe Efstathiou, G., Bernstein, G., Katz, N., \,\,\, Tyson, J.A., 
      Guhathakurta, P., 1991, \apj 380, L47
\refe Eke, V.R., Cole, S., Frenk, C.S., 1996, \mn 282, 263
\refe Giavalisco, M., Steidel, C.C., Adelberger, K.L., Dickinson, M.E., 
      Pettini, M., Kellogg, M., 1998, \apj 503, 543
\refe Governato, F., Baugh, C.M., Frenk, C.S., Cole, S., Lacey, C.G., 
      Quinn, T., Stadel, J., 1998, Nature, 392, 359
\refe Groth, E.J., Peebles, P.J.E., 1977, \apj 217, 38
\refe Hamilton, A.J.S., Kumar, P., Lu, E., Matthews, A., 1991, \apj 374, L1
\refe Heyl, J.S., Cole, S., Frenk, C.S., Navarro, J.F., 1995, \mn 274, 755
\refe Hudon, J.D., Lilly, S.J., 1996, \apj 469, 519
\refe Infante, L., Pritchet, C.J., 1995, \apj 439, 565
\refe Jenkins,A.,  Frenk, C.S.,  Pearce, F.R.,  Thomas, P.A., Colberg, J.M., 
      White, S.D.M.,  Couchman, H.M.P., Peacock, J.A.,  Efstathiou,G.,  
      Nelson, A.H., 1998, \apj, 499, 20  
\refe Jing, Y.P., 1998, \apj 503, L9
\refe Kauffmann, G., Nusser, A., Steinmetz, M., 1997, \mn 286, 795
\refe Kauffmann, G., White, S.D.M., Guiderdoni, B., 1993, \mn 264, 201
\refe Kauffmann, G., Colberg, J.M., Diaferio, A., White, S.D.M., 1999, \mn
      303, 188.
\refe Kauffmann, G., Colberg, J.M., Diaferio, A., White, S.D.M., 1998, \mn
      submitted, astro-ph/9809168.
\refe Lacey, C.G., Cole, S. 1993 \mn 262, 627
\refe Lacey, C.G., Guiderdoni, B., Rocca-Volmerange, B., Silk, J.,
      1993, \apj 402, 15
\refe LeF\`{e}vre, O., Hudon, D., Lilly, S.J., Crampton, D., Tresse, L., 
      1996, \apj 461, 534.
\refe Limber, D.N., 1954, \apj 119, 655
\refe Matarrese, S., Coles, P., Lucchin, F., Moscardini, L., 1997, \mn
      286, 115
\refe Mo, H.J., Jing, Y.P., White, S.D.M., 1996, \mn 282, 1096
\refe Mo, H.J., White, S.D.M., 1996, \mn 282, 347
\refe Mo, H.J., Mao, S., White, S.D.M., 1999, \mn, 304, 175
\refe Moscardini, L., Coles, P., Lucchin, F., Matarrese, S, \mn, 299, 95
\refe Peacock, J.A., Dodds, S.J., 1996, \mn 280, L19
\refe Peebles, P.J.E., 1980, Large Scale Structure in the Universe 
      Princeton.
\refe Postman, M., Lauer, T.R., Szapudi, I., Oegerle, W., 1998, \apj 506, 33
\refe Roche, N., Shanks, T., Metcalfe, N., Fong, R., 1993, \mn 263, 360
\refe Somerville, R.S, Primack, J.R., 1998, astro-ph/9802268
\refe Steidel, C.C., Giavalisco, M., Pettini, M., Dickinson, M., 
      Adelberger, K.L., 1996, \apj 462, L17      
\refe White, S.D.M., Frenk, C.S., 1991, \apj 379, 52
\end{document}